\documentclass[10pt,letterpaper]{article}
\usepackage[top=0.85in,left=2.75in,footskip=0.75in]{geometry}

\usepackage{changepage}

\usepackage[utf8]{inputenc}

\usepackage{textcomp,marvosym}

\usepackage{amsmath,amssymb}

\usepackage{cite}
\usepackage{tabularx}

\usepackage{nameref,hyperref}

\usepackage{booktabs} % Horizontal rules in tables
\usepackage[right]{lineno}

\usepackage{microtype}
\DisableLigatures[f]{encoding = *, family = * }

\usepackage{rotating}

\raggedright
\usepackage[aboveskip=1pt,labelfont=bf,labelsep=period,justification=raggedright,singlelinecheck=off]{caption}

\makeatletter
\renewcommand{\@biblabel}[1]{\quad#1.}
\makeatother

\date{}

\usepackage{lastpage,fancyhdr,graphicx}
\usepackage{epstopdf}
\fancyheadoffset[L]{2.25in}
\fancyfootoffset[L]{2.25in}
\begin{document}
	
	%\preprint{}
	
	\title{}
	
\vspace*{0.35in}

% Please capitalize all terms in the title except conjunctions, prepositions, and articles.
\begin{flushleft}
{\Large
\textbf\newline{\bf Phone-based Metric as a Predictor for Basic Personality Traits}
}
\newline
\\
Bjarke Mønsted,\\
Anders Mollgaard,\\
Joachim Mathiesen\textsuperscript{*}\\
\bigskip
University of Copenhagen, Niels Bohr Institute , 2100 Copenhagen, Denmark\\
\bigskip
% Use the asterisk to denote corresponding authorship and provide email address in note below.
* mathies@nbi.dk

\end{flushleft}

\section*{Abstract}
Basic personality traits are typically assessed through questionnaires. Here we consider phone-based metrics as a way to asses personality traits. We use data from smartphones with custom data-collection software distributed to 730 individuals. The data includes information about location, physical motion, face-to-face contacts, online social network friends, text messages and calls. The data is further complemented by questionnaire-based data on basic personality traits. From the phone-based metrics, we define a set of behavioural variables, which we use in a prediction of basic personality traits. We find that predominantly, the Big Five personality traits extraversion and, to some degree, neuroticism are strongly expressed in our data. As an alternative to the Big Five, we investigate whether other linear combinations of the 44 questions underlying the Big Five Inventory are more predictable. In a tertile classification problem, basic dimensionality reduction techniques, such as independent component analysis, increase the predictability relative to the baseline from $11\%$ to $23\%$. Finally, from a supervised linear classifier, we were able to further improve this predictability to $33\%$. In all cases, the most predictable projections had an overweight of the questions related to extraversion and neuroticism. In addition, our findings indicate that the score system underlying the Big Five Inventory disregards a part of the information available in the 44 questions.

\section{Introduction}
Over the last decades, new data collection methods have provided new opportunities for research on human behavior. Online social networks or personal mobile devices do not only provide real-time data for studies on human activity and interaction, but can also serve as an external validation of e.g. more classical questionnaire- or interview-based studies. For example, the predictability of basic personality traits from smart-phone usage is currently an active area of research \cite{Sekara2014, demontjoye2013, DeOliveira2011, LiKamWa2011, Verkasalo2010, Chittaranjan2011, Chittaranjan2011a, Crandall2010, MJW}. 

Much of this research is devoted to predicting the Big Five personality traits\cite{Digman1990a}, openness (O), conscientiousness (C), extraversion (E), agreeableness (A) and neuroticism (N), commonly called the five factor model and abbreviated as OCEAN. %The Big Five provides an established taxonomy for personality research \cite{john2008paradigm}.
The five traits are estimated by letting individuals fill a questionnaire with 44 items. Each item consists of a statement and is answered by expressing how much one agrees with this statement on a discrete scale from 1 to 5. The personality traits then follow from a pre-determined linear combination of the 44 answers. Notwithstanding that phone usage predominantly reflects social interaction, studies report surprisingly accurate predictions of all the Big Five personality traits. For example, one would by default not expect a trait such as conscientiousness to be strongly expressed in phone usage.

\section{Methods}
We use questionnaire-based data on the personality traits together with phone based-data from $730$ freshman students starting in the year 2013 at the Technical University of Denmark. The phone-based data has been collected over a period of 24 months by custom software installed on smartphones given to the participants of the study \cite{mollgaard2016,Stopczynski2014}. The data consists of telecommunication logs (phone calls, text messages), online social networks (Facebook connections and interactions), and networks based on physical proximity. The physical proximity is measured through the Bluetooth signal strength, and can be used to monitor face-to-face contacts~\cite{Sekara2014}. From the GPS, the data also includes information on the geo-spatial mobility. Out of the 730 participants, we only included data from participants, which, we believe, used the phone as a primary device. This implied discarding data from users with less than 10 texts, 5 calls or 100 GPS points, as well as users with no Facebook friends. These requirements reduced the number of participants in our study to 636.

Similar to \cite{demontjoye2013}, we extract a range of features from the smartphone data, which we will use in a machine learning scheme to predict the personality traits. Below we provide details about the features and about how they are used in the prediction of the individual traits. Table~\ref{tab:feats} presents a list of all the features we consider. We finally provide a brief description of the various dimensionality reduction techniques we apply to the 44 questions underlying the Big Five.

\emph{Feature Extraction}. The first category of features that we extract consists of basic measures of calls and texting. For each user, we compute the median and standard deviation of the inter-event time between phone calls, text messages, and combinations thereof. For each of the three interaction forms, we also compute the entropy $S_u$ defined by
\begin{equation}
	S_u = \sum_c \frac{n_c}{n_t} \log_2\frac{n_c}{n_t}, \label{eqn:entropy_thingy}
\end{equation}
where the index $c$ runs over each unique number that the user has contacted, $n_c$ denotes the number of interactions with contact $c$, and $n_t$  the total number of interactions. The entropy is a general measure of the spread of the interactions. Users with low entropy tend to have most of their interactions with a few contacts. We further determine the percentage of a user's calls, which are categorised as outgoing, as well as the total number of contacts, their ratio to the number of interactions, and the ratio of calls and texts that a user has responded to within an hour of receiving them, and finally the fraction of calls the took place during the night.

A number of quantities based on location data are also computed. We extract the median and standard deviation of the users' daily distance travelled, their daily radius of gyration (the radius of the smallest circle enclosing all coordinates visited by the user on each day) and the entropy of the time spend in various locations by the user.
We compute this by first finding GPS points at which the user was stationary, and then looking for clusters within the stationary points. We considered a point stationary if the user's mean velocity didn't exceed 0.5 m/s in the period between two consecutive points. As the uncertainty on civilian GPS locations can be up to 100m\cite{gps_precision}, we consider only data points taken at least 500 seconds apart. We then use the DBSCAN algorithm\cite{Ester1996} to identify clusters and compute the entropy of the points in those clusters similarly to \eqref{eqn:entropy_thingy}. Finally, we estimate the fraction of time a user spends at home, where home is assumed to be the place where a user spend most of their weeknights.

Another category of features aim to quantify the degree to which a user's behavior follows a temporal pattern. Using call/text data to describe this with an example, we count the number of call/text events for a given user in bins of 6 hours. We then fit an autoregressive series to best predict the activity in time bin $X_t$ from previous activity on the form
\begin{equation}
	X_t = \mu + \epsilon_t + \sum_{i+1}^p \varphi_i X_{t-i}, \label{eqn:time_series}
\end{equation}
where $\mu$ is the mean activity and $\epsilon_t$ is a noise term.

We finally extract a range of features concerning a user's social contacts. This includes their number of Facebook friends and the fraction of the time users spend in the proximity of other participants in the study. This is estimated from repeated automatic scans by the Bluetooth ports. The entropy of the proximity is also calculated similarly to Eq.~\eqref{eqn:entropy_thingy}, as well as the time series parameters as described in Eq.~\eqref{eqn:time_series}.

\begin{table}
\centering
%[H] add [H] placement to break table across pages
	\caption{{\bf Features included in the classifiers.} Table of the features (first column) included in the classifiers for each of the Big Five traits (second column) abbreviated openness (O), conscientiousness (C), extraversion (E), agreeableness (A) and neuroticism (N). In the classifier, we include only the features with the highest linear correlation with the trait scores. Starting from the features with the highest correlation, we determine the number of features to include from a grid search. The grid search procedure is described in the Methods section.}\label{tab:feats}
% Fjern variabelnavne hvis de ikke publiseres!!!
\begin{tabularx}{\textwidth}{ll}
\hline\hline
\textbf{Feature}  & \textbf{Trait(s)} \\
\hline
Bluetooth autocovariance coefficient  &  O\\ 
Contact entropy using 24-hour bins  &  N\\ 
Call duration (median) &  N\\ 
Call duration (standard deviation)  &  N\\ 
Call inter-event time (standard deviation)  &  OE\\ 
Percent of a user's calls initiated by themselves  &  A\\ 
Call/text contact-interaction ratio  &  A\\ 
Call/text inter-event time median &  N\\ 
Combined call/text inter-event time (standard deviation) &  CEAN\\ 
Ingoing call/text autoregressive series coefficient 13  &  O\\ 
Ingoing call/text autoregressive series coefficient 4 &  C\\ 
Number of contacts during the first three months &  OE\\ 
Number of call/text events &  C\\ 
Number of texts &  C\\ 
Number of Facebook friends &  OE\\ 
Outgoing call/text AR series coefficient 2 &  C\\ 
Outgoing call/text AR series coefficient 4 &  C\\ 
Text contact/interaction ratio &  A\\ 
Text inter-event time (median) &  N\\ 
Text inter-event time (standard deviation)  &  CEAN\\ 
Median text response time &  O\\ 
Fraction of texts that were outgoing &  A\\ 
fraction of texts responded to within an hour &  CAN\\
\hline\hline
\end{tabularx}
\end{table}

\subsection{Classification}
We divide the scores on each of the five personality traits in tertiles, similar to \cite{demontjoye2013}, and assign a label of 0, 1 or 2 specifying whether they score low, medium, or high on that trait. Our model of choice for predicting these labels $\mathbf{Y}$ from the feature vectors $\mathbf{X}$ is a support vector machine (SVM) using a radial basis function (RBF) kernel\cite{hearst1998}. This model requires that two hyperparameters are fixed - a misclassification cost $C$ and a sharpness $\gamma$ of the Gaussian basis functions. Additionally, we find the number of features to include in the classifier by finding the linear correlation of each feature to the target label and including only the $n$ features with the strongest correlation. We fix these three parameters by performing a grid search over the parameter space and running 100 stratified 5-fold validation procedures (randomly splitting the data in 5 test sets of $20\%$ of the data, using the remaining $80\%$ for training) for each combination, keeping the combination of hyperparameters giving the highest mean classification score. We compared the results with a linear SVM, which showed that using an RBF kernel only gave a very slight improvement over a linear classifier. The optimal values for $C$ and $\gamma$ are shown in Table \ref{tab:hyperpars}, and the features used to classify each big five trait are shown in Table \ref{tab:feats}.

\begin{table}%
	\centering
	\caption{{\bf Choice of hyperparameters}. The values for the hyperparameters $C$ and $\gamma$ which gave the best prediction in the grid search.}
	\label{tab:hyperpars}
	\begin{tabular}{|c|ccccc|}
		\hline
		& Openness & Conscient. & Extrav. & Agreeable. & Neuroticism\\ 
\hline
		$C$ & $0.8$ & $0.8$ & $1.0$ & $42.0$ & $1.0$ \\
		$\gamma$ & $0.2$ & $2.0$ & $0.05$ & $0.75$ & $1.0$ \\
\hline
		\end{tabular}
\end{table}

We further test the validity of the scoring system applied to the 44 items of the Big Five questionnaire. We do this by considering alternative dimensionality reduction techniques. Specifically, we use principal component analysis (PCA), independent component analysis (ICA), factor analysis (FA), and supervised dimensionality reduction (SDR), keeping only the five leading components of each technique. The supervised dimension reduction technique applied here finds the one dimensional projection of the data that has the lowest $R^2$ value, when training a linear model. The procedure is continued with the additional constraint that the new projections should be orthogonal to all previous projections, such that the result is a low dimensional space specified by an orthogonal basis. The constrained optimization is performed numerically on the training set and then applied to the test set in order to avoid overfitting. See the Supporting Information S1: SDR.

\section{Results}
The effectiveness of our classification is measured in terms of the mean \textit{relative improvement} over a baseline classifier
\begin{equation}\label{RI}
	S = \left\langle \frac{f_\text{classifier}}{f_\text{baseline}} - 1 \right\rangle,
\end{equation}
where $f$ denotes the fraction of correct classifications. We use the \textit{Friends and Family} dataset\cite{Aharony2011} to test the performance of our machine learning scheme. The publicly available part of the dataset%
\footnote{Available at http://realitycommons.media.mit.edu/friendsdataset.html} 
consists of data from 52 participants, 38 of which have sufficient call and location data for our analysis (the remaining 14 participants has either fewer than 10 texts, fewer than 5 calls, fewer than 100 GPS points or did not have data on all of the 44 questions need to compute their Big Five traits).

We reach a mean relative improvement of $0.31 \pm 0.06$ over $10^4$ bootstrap samples, where the error of $0.06$ is the standard deviation of the results. This result is in agreement with the result reported in \cite{demontjoye2013} of $0.42$,  which utilize a slightly different data set. The results are significantly above what is reported in other studies\cite{Chittaranjan2013}. Here, we analyze further the predictability of personality by using an extensive data set. We also consider the impact of predictability by different dimensionality reductions of the 44 questions in the Big Five Inventory.

%\begin{figure}
%	\centering
%	\includegraphics[width = 0.7\textwidth]{figures/ff_histogram.pdf}
%	\caption{Relative improvement of our SVM classifier over a baseline classifier which simply guesses on the most commonly occurring class in the training set. Over $10^4$ runs each of which partitioned the data randomly into training and test sets. These runs had a mean improvement over baseline of $0.31$ and a standard deviation of $0.06$.}
%	\label{fig_ff}
%\end{figure}

\begin{table}
\centering
%[H] add [H] placement to break table across pages
\caption{{\bf Performance of the classifier}. Comparison of the performance of our classifier on each of the Big Five traits in our dataset ($n=636$) with the Friends and Family dataset ($n=38$).}
\label{tab:results}
\begin{tabular}{lllllll}
\hline\hline
Trait & Openness & Conscient. & Extrav. & Agreeable. & Neuroticism & Mean \\ 
\hline
FF & $35.3 \pm 11.5$ & $16.6 \pm 13.9$ & $49.3 \pm 12.7$ & $17.6 \pm 15.1$ & $37.1 \pm 17.9$ & $31.2 \pm 6.4$\\ 
SFP & $6.2 \pm 2.8$ & $-2.4 \pm 2.4$ & $36.7 \pm 1.1$ & $8.4 \pm 3.2$ & $5.5 \pm 2.2$ & $10.9 \pm 1.1$\\
\hline\hline
\end{tabular}
\end{table}

%Here we consider a much more extensive data set based on 636 active participants. 

\begin{figure}
	\centering
	\includegraphics[width = 0.7\textwidth]{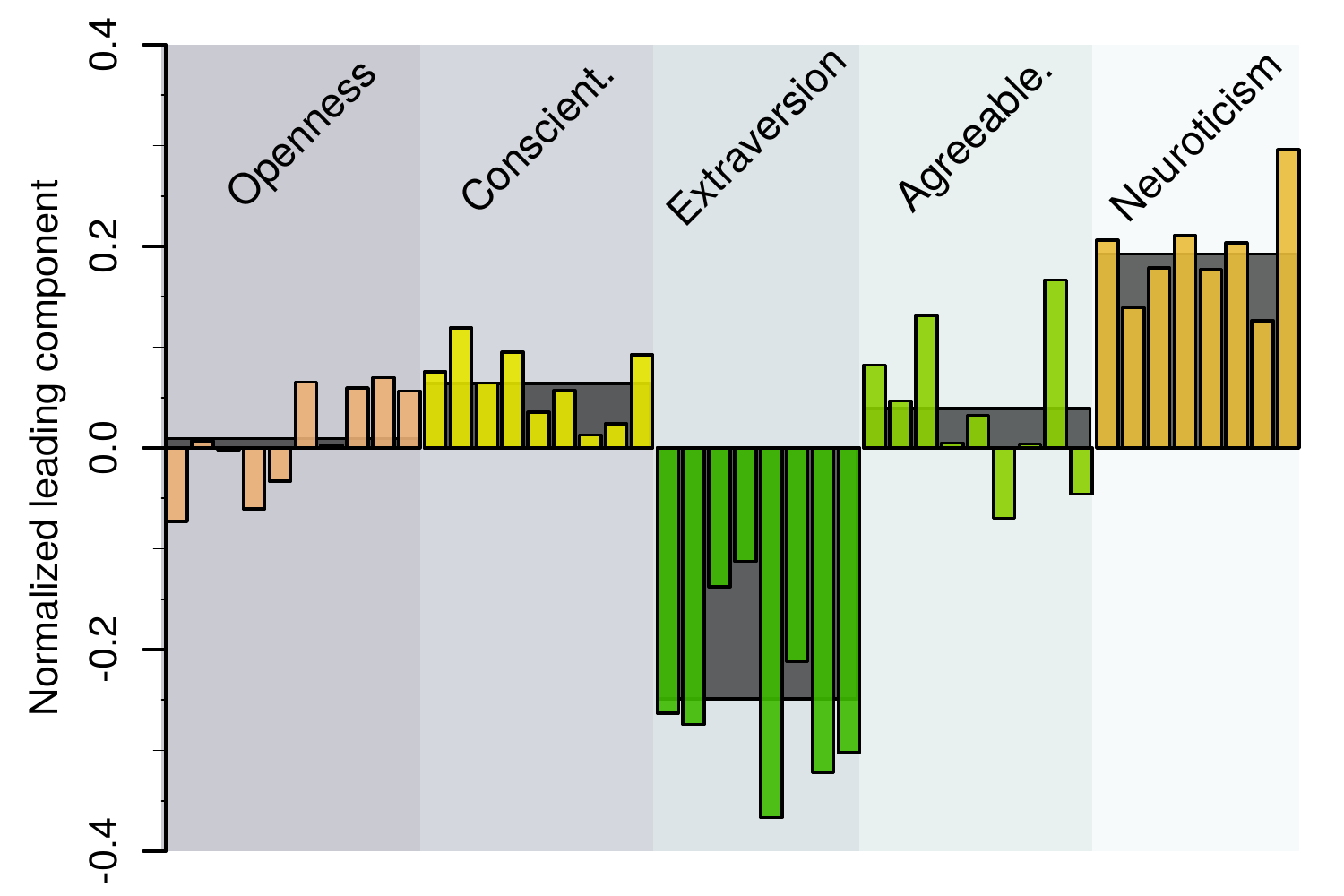}
	\caption{{\bf The ICA component with the highest predictability}. The 44 entries are grouped according to which big five trait the corresponding question is associated with. The wider bars behind show the mean value of each group of questions, thus denoting how strongly associated the ICA component is with each of the five traits.}
	\label{fig:coeffs}
\end{figure}

Interestingly and in contrast to the \textit{Friends and Family} data set, we only achieve a relative improvement to the baseline of $S=0.11 \pm 0.01$ for our larger data set.
\begin{figure}
	\centering
	\includegraphics[width = 0.7\textwidth]{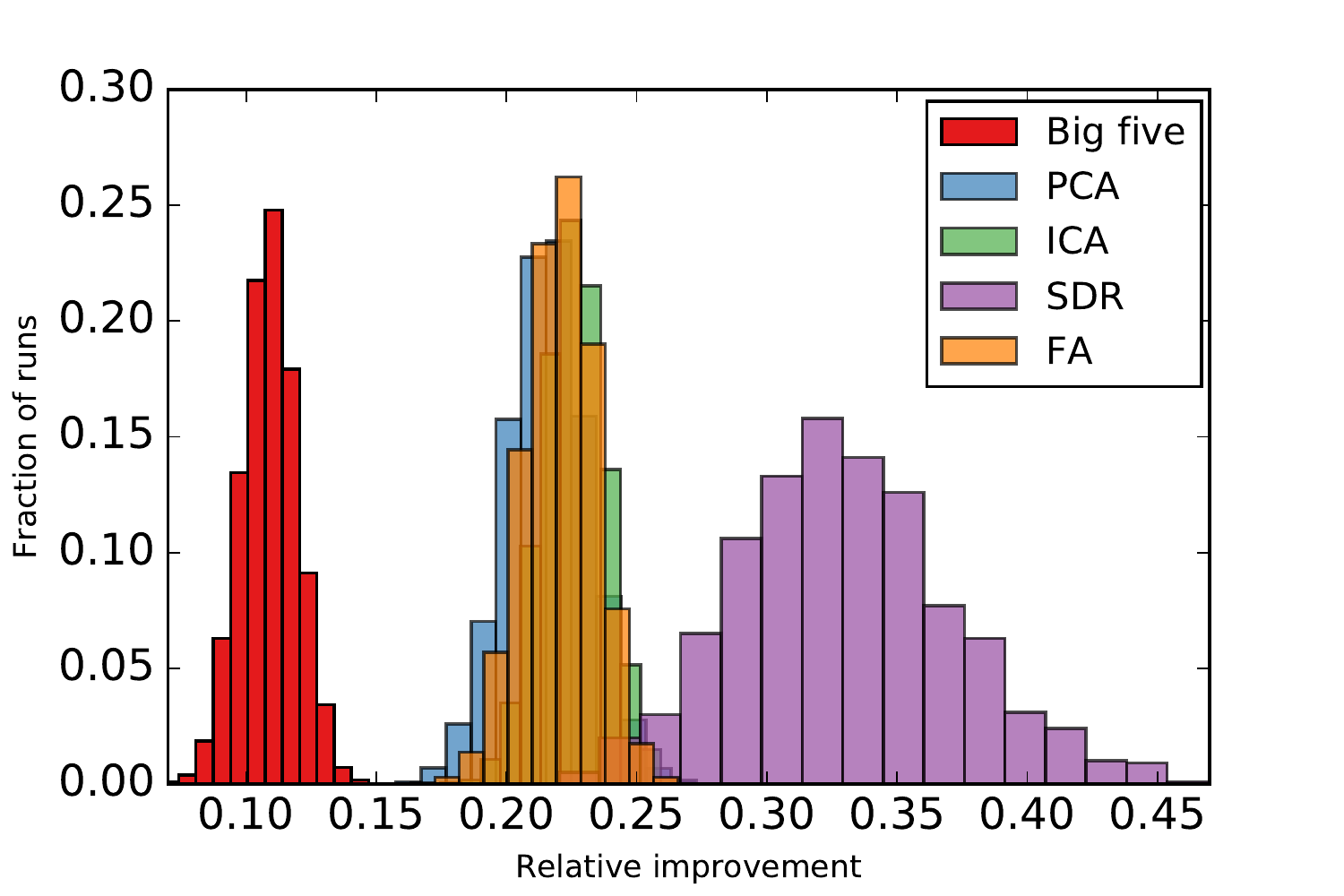}
	\caption{{\bf Comparison of the predictability of the Big Five personality for different dimensionality reduction techniques}. The plot shows the distribution of relative improvements for $10^4$ cross validation runs using various projections of the data. Apart from the Big Five projections (with a mean predictability increase of $0.11\pm0.01$), the unsupervised dimensional reduction techniques include principal component analysis (PCA, $0.22\pm0.02$), independent component analysis (ICA, $0.23\pm0.01$) and factor analysis (FA, $0.22\pm0.01$). The supervised dimensionality reduction (SDR) technique reached an improvement of $0.33\pm0.04$.}
	\label{fig:sfp_data}
\end{figure}
Insofar as a person's true personality can be assumed to be reflected in their phone behaviour, it is problematic that the Big Five Inventory retains less predictive power from the original 44 questions used to compute it, than any of the other dimensionality reduction techniques investigated. As can be seen in Table \ref{tab:results}, extraversion can be predicted with much greater accuracy. This is perhaps not surprising given that our features come from smartphones, which to a large extent is used to establish and maintain social ties. To investigate this further, we examined the components of the projection vectors used in each dimensionality reduction technique. In all cases, the projection retaining the greatest predictability was strongly associated with extraversion and in many cases also with neuroticism. For example, Fig.~\ref{fig:coeffs} shows the entries of the ICA vector whose projection had the greatest predictability.

\section{Concluding Remarks}
In general, only extraversion and to some degree neuroticism can be predicted by smartphone usage patterns, whereas the remaining personality traits of the Big Five  have limited predictability. This is further supported by the fact that when trying several different dimensionality reduction techniques, we repeatedly found that the directions along which a projection retained the most predictability had much stronger associations with extraversion and neuroticism than with the remaining Big Five traits. We also found that dimensionality reduction schemes alternative to the common Big Five scoring system could significantly increase the predictability.

\section{Data Availability and Ethics Statement}
Data are part of larger study "Social Fabric" involving researchers at the Technical University of Denmark and University of Copenhagen. Due to privacy consideration regarding subjects in our dataset, including European Union regulations and Danish Data Protection Agency rules, we cannot make all data used here publicly available. The data contains detailed information on mobility and daily habits at a high spatio-temporal resolution. We understand and appreciate the need for transparency in research and are ready to make the data available to researchers who meet the criteria for access to confidential data, sign a confidentiality agreement, and agree to work under our supervision in Copenhagen. The "Social Fabric" study was reviewed and approved by the appropriate Danish authority, the Danish Data Protection Agency (Reference number: 2012-41-0664). The Data Protection Agency guarantees that the project abides by Danish law and also considers potential ethical implications. All subjects in the study gave written informed consent.
%\appendix
%\section{}

%\begin{acknowledgments}
%\end{acknowledgments}

\section*{Supporting Information} 
\subsection*{Supervised dimensionality reduction}

In this section we explain in greater detail the supervised dimensional reduction technique applied in the paper. The goal is to find the projections of the 44 questions, which we can predict the best. 

The questionnaire data is represented by a matrix, $y_{ij}$, where $i$ denotes a person and $j$ denotes a question. Similarly, we have a matrix describing smartphone behavior, $x_{ij}$, where $i$ denotes a person and $j$ denotes the behavioural variable. The projection vector, $p_j$, is 44 dimensional and has unit length

\begin{equation}
	1 = \sum_j p_j^2.
\end{equation}
It reduces the information in the 44 questions to a single number through an inner product

\begin{equation}
	y_i^{(p)} = \sum_j y_{ij} p_j.
\end{equation}
We introduce a linear model to estimate this value based on the behavioural variables

\begin{align*}
	y_i^{(p)} &= \sum_j x_{ij} \alpha_j + \epsilon_i,
\end{align*}
where $\epsilon_i$ is the error of the model estimate for person $i$. We aim to train the projection vector, $p_j$, and the linear model parameters, $\alpha_j$, such that the coefficient of determination, $R^2$, is as large as possible. The coefficient is defined as

\begin{equation}
	R^2 = 1 - SS_\mathrm{res} / SS_\mathrm{tot},
\end{equation}
where 

\begin{equation}
	SS_\mathrm{res} = \sum_i \epsilon_i^2,
\end{equation}
and

\begin{equation}
	SS_\mathrm{tot} = \sum_i \left( y_i^{(p)} - \bar{y}^{(p)} \right)^2,
\end{equation}
with $\bar{y}^{(p)}$ the average projection over the persons. The training is performed iteratively in two steps. First, we fix the projection vector and optimize for the parameters of the linear model. Then we fix the parameters and optimize for the projection vector. The optimization step is performed using Sequential Least SQuares Programming (SLSQP) with the projection vector constrained to unit length. The training converges consistently irrespective of the initialization of the projection vector.

We may then look for the best projection in the 43 dimensional space orthogonal to our first projection. This can either be done by mapping on to these 43 dimensions or simply adding an orthogonality constraint to the optimization. This procedure may be repeated until a satisfying number of projections is obtained.

We have a final note regarding over training. Let us start by counting the number of free parameters in the training step. If the dimension of $y$ is $N$ and the dimension of $x$ is $M$, then the number of free parameters is $M + N$, since the linear model has an extra parameter for offset, which is cancelled by the unit length constraint on the projection vector. For a data set of size $S$, we need $S \gg M + N$ for proper training. In other words, if too many features of $x$ are included in the SDR scheme, fitting to noise will take place, thereby resulting in worse performance when applying the classifier to a test set. To avoid this over fitting effect, we implement the following procedure to determine the optimal features of $x$ to include. First we partition the data into five test, and training, sets consisting of 80\% and 20\% of the data, respectively. Within each training set we find the correlation between the features of $x$ and each of the 44 features of $y$. For each feature, we compute the product of the p-values corresponding to those correlations, obtaining a value between 0 and 1, where a value of 1 is interpreted as the feature being unrelated to $y$ and lower values indicating stronger associations. We then rank the features according to these values, and keep $n$ best features for the classification task. We find that $n=8$ performs the best, since over fitting takes over for larger $n$, and we therefore use these 8 features for the supervised dimension reduction.

\end{document}